\documentclass[twocolumn,showpacs,preprintnumbers,nofootinbib,prd,superscriptaddress,groupedaddress,10pt]{revtex4-1}

\usepackage{graphicx,amssymb,amsmath,amsthm,amsfonts,epsfig,epsf}
\usepackage[linktocpage]{hyperref}
\usepackage[usenames]{color}
\usepackage{epstopdf}

\usepackage{bm}
\usepackage{dcolumn}
\usepackage[utf8]{inputenc}
\usepackage{latexsym}
\usepackage{rotating}
\usepackage{hyperref}
\usepackage{color}
\usepackage{longtable}
\usepackage{enumerate}
\usepackage{tensor}
\usepackage{url}
\newcommand{\ben}{\begin{enumerate}}
\newcommand{\een}{\end{enumerate}}

\def\be{\begin{equation}}
\def\ee{\end{equation}}
\def\bea{\begin{eqnarray}}
\def\eea{\end{eqnarray}}

\newcommand{\beq}{\begin{eqnarray}}
\newcommand{\eeq}{\end{eqnarray}} 
\newcommand{\ba}{\begin{align}}
\newcommand{\ea}{\end{align}}

\begin{document}

\title{Viewpoint: The First Sounds of Merging Black Holes}

\author{Emanuele Berti$^{1,2}$}
\affiliation{${^1}$ Department of Physics and Astronomy, The University of Miss
issippi, University, MS 38677, USA}
\affiliation{${^2}$ CENTRA, Departamento de F\'{\i}sica, Instituto Superior T\'ecnico -- IST, Universidade de Lisboa -- UL,
Avenida Rovisco Pais 1, 1049 Lisboa, Portugal}

\begin{abstract}
Gravitational waves emitted by the merger of two black holes have been detected, setting the course for a new era of observational astrophysics.
\end{abstract}



\maketitle


For decades, scientists have hoped they could ``listen in'' on violent astrophysical events by detecting their emission of gravitational waves. The waves, which can be described as oscillating distortions in the geometry of spacetime, were first predicted to exist by Einstein in 1916, but they have never been observed directly. Now, in an extraordinary paper, scientists report that they have detected the waves at the Laser Interferometer Gravitational-wave Observatory (LIGO)~\cite{Abbott:2016blz}. From an analysis of the signal, researchers from LIGO in the US, and their collaborators from the Virgo interferometer in Italy, infer that the gravitational waves were produced by the inspiral and merger of two black holes (Fig.~1), each with a mass that is more than 25 times greater than that of our Sun. Their finding provides the first observational evidence that black hole binary systems can form and merge in the Universe.

\begin{figure}[bht] \includegraphics[scale=0.3,clip=true,angle=0]{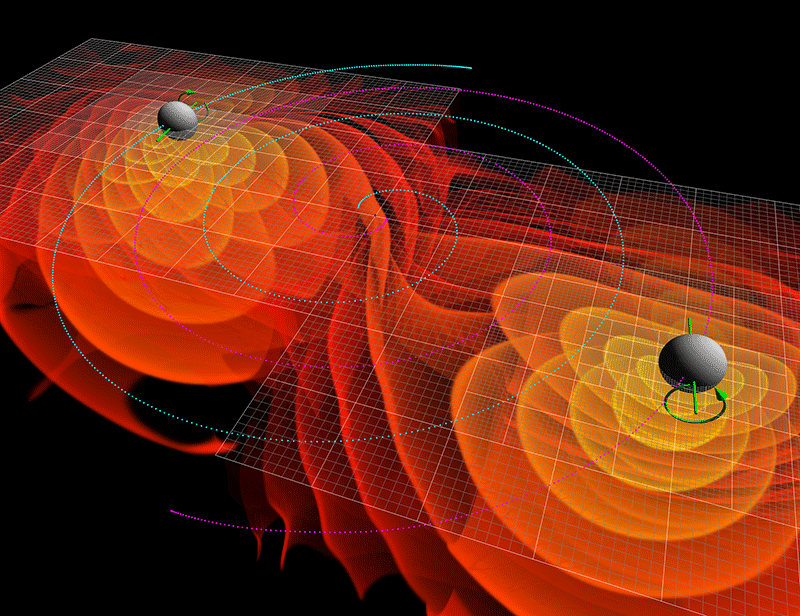} \caption{Numerical simulations of the gravitational waves emitted by the inspiral and merger of two black holes. The colored contours around each black hole represent the amplitude of the gravitational radiation; the blue lines represent the orbits of the black holes and the green arrows represent their spins. [Credit: C. Henze/NASA Ames Research Center]}
\label{fig1}
\end{figure}

Gravitational waves are produced by moving masses, and like electromagnetic waves, they travel at the speed of light. As they travel, the waves squash and stretch spacetime in the plane perpendicular to their direction of propagation (see inset, Fig.~2). Detecting them, however, is exceptionally hard because they induce very small distortions: even the strongest gravitational waves from astrophysical events are only expected to produce relative length variations of order $10^{-21}$.

\begin{figure*} \includegraphics[scale=0.6,clip=true,angle=0]{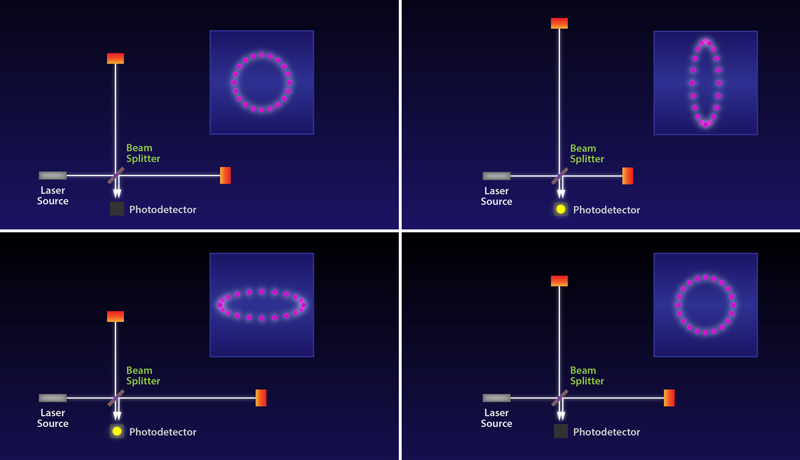} \caption{A schematic depiction of LIGO's interferometric gravitational wave detector. Light from a laser is split in two by a beam splitter; one half travels down the vertical arm of the interferometer, the other half travels down the horizontal arm. The detector is designed so that in the absence of gravitational waves (top left) the light takes the same time to travel back and forth along the two arms and interferes destructively at the photodetector, producing no signal. As the wave passes (moving clockwise from top right) the travel times for the lasers change, and a signal appears in the photodetector. (The actual distortions are extremely small, but are exaggerated here for easier viewing.) Inset: The elongations in a ring of particles show the effects of a gravitational wave on spacetime. [Credit: APS/Alan Stonebraker]}
\label{fig2}
\end{figure*}

``Advanced'' LIGO, as the recently upgraded version of the experiment is called, consists of two detectors, one in Hanford, Washington, and one in Livingston, Louisiana. Each detector is a Michelson interferometer, consisting of two 4-km-long optical cavities, or ``arms,'' that are arranged in an L shape. The interferometer is designed so that, in the absence of gravitational waves, laser beams traveling in the two arms arrive at a photodetector exactly $180^\circ$ out of phase, yielding no signal. A gravitational wave propagating perpendicular to the detector plane disrupts this perfect destructive interference. During its first half-cycle, the wave will lengthen one arm and shorten the other; during its second half-cycle, these changes are reversed (see Fig.~2). These length variations alter the phase difference between the laser beams, allowing optical power -- a signal -- to reach the photodetector. With two such interferometers, LIGO can rule out spurious signals (from, say, a local seismic wave) that appear in one detector but not in the other.

LIGO's sensitivity is exceptional: it can detect length differences between the arms that are smaller than the size of an atomic nucleus. The biggest challenge for LIGO is detector noise, primarily from seismic waves, thermal motion, and photon shot noise. These disturbances can easily mask the small signal expected from gravitational waves. The upgrade, completed in 2015, improved the detector's sensitivity by a factor of 3--5  for waves in the 100--300~Hz frequency band and by more than a factor of 10 below 60~Hz. These improvements have enhanced the detector's sensitivity to more distant sources and were crucial to the discovery of gravitational waves.

\begin{figure*} \includegraphics[scale=0.6,clip=true,angle=0]{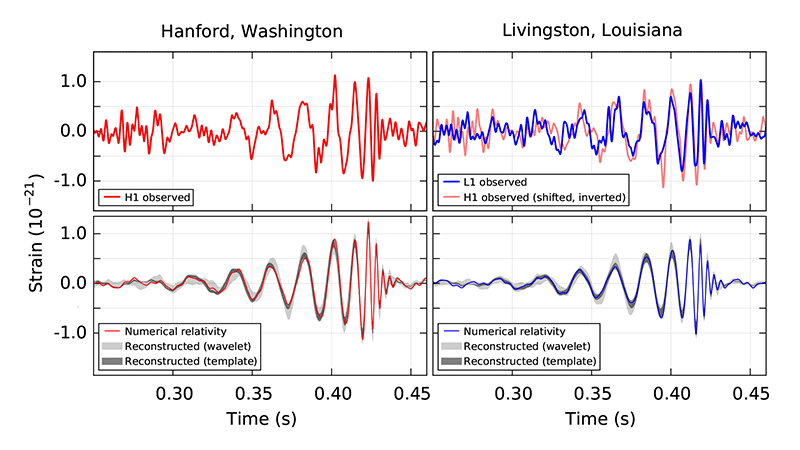} \caption{On September 14, 2015, similar signals were observed in both of LIGO's interferometers. The top panels show the measured signal in the Hanford (top left) and Livingston (top right) detectors. The bottom panels show the expected signal produced by the merger of two black holes, based on numerical simulations. [Credit: B. P. Abbott et al. [1].]}
\label{fig3}
\end{figure*}

On September 14, 2015, within the first two days of Advanced LIGO's operation, the researchers detected a signal so strong that it could be seen by eye (Fig.~3). The most intense portion of the signal lasted for about 0.2~s and was observed in both detectors, with a combined signal-to-noise ratio of 24. Fittingly, this first gravitational wave signal, dubbed GW150914, arrived less than two months before the 100--year anniversary of the publication of Einstein's general relativity theory.

Up until a few decades ago, detecting gravitational waves was considered an impossible task. In fact, in the 1950s, physicists were still heatedly debating whether the waves were actual physical entities and whether they could carry energy. The turning point was a 1957 conference in Chapel Hill, North Carolina \cite{Kennefick:2007zz,Saulson:2010zz}. There, the theorist Felix Pirani pointed out a connection between Newton's second law and the equation of geodesic deviation, which describes the effect of tidal forces in general relativity. This connection allowed him to show that the relative accelerations of neighboring particles in the presence of a gravitational wave provide a physically meaningful -- and measurable -- way to observe it. Sadly, Pirani, who laid the groundwork for our modern thinking about gravitational waves and how to detect them, passed away on December 31, 2015, just weeks before the LIGO scientists announced their discovery.

Other prominent physicists at the meeting, including Joseph Weber, Richard Feynman, and Hermann Bondi, were instrumental in pushing Pirani's ideas forward. Feynman and Bondi, in particular, developed Pirani's observation into what is now known as the ``sticky bead'' thought experiment. They argued that if beads sliding on a sticky rod accelerated under the effect of a passing gravitational wave, then they must surely also transfer heat to the rod by friction. This heat transfer is proof that gravitational waves must indeed carry energy, and are therefore, in principle, detectable.

Interest in carrying out such experiments wasn't immediate. As Pirani noted in his 1964 lectures on gravitational radiation~\cite{Pirani1965}, Weber thought that meaningful laboratory experiments were ``impossible by several orders of magnitude.'' At about the same time, William Fowler (the future Nobel laureate) suggested that a large fraction of the energy emitted by so-called massive double quasars -- what we now know as black hole binaries -- might be in the form of gravitational radiation. Pirani, however, felt that the direct observation of gravitational waves was not ``necessary or sufficient'' to justify a corresponding theory, arguing that unless physicists figured out a way to quantize gravity, such a theory would not ``have much to do with physics''~\cite{Pirani1965}.

What galvanized the field was a 1969 paper from Weber, who claimed he had detected gravitational radiation with a resonant bar detector (see \href{http://physics.aps.org/story/v16/st19}{\underline{22 December 2005 Focus}} story). The finding was controversial -- physicists could not duplicate it and by the mid--1970s, most agreed that Weber had likely been incorrect. However, a few years later a young professor at the Massachusetts Institute of Technology named Rainer Weiss was preparing for his course on relativity when he came across a proposal by Pirani for detecting gravitational waves. Pirani had suggested using light signals to see the variations in the positions of neighboring particles when a wave passed. His idea, with one key modification, led to the genesis of LIGO: rather than using the timing of short light pulses, Weiss proposed to make phase measurements in a Michelson interferometer~\cite{Weiss1972}. Ronald Drever, Kip Thorne, and many others made crucial contributions to developing this idea into what LIGO is today. (See Ref.~\cite{Kennefick:2007zz} for a historical account.)

Now, what was once considered ``impossible by several orders of magnitude'' is a reality. To confirm the gravitational-wave nature of their signal, the researchers used two different data analysis methods. The first was to determine whether the excess power in the photodetector could be caused by a signal, given their best estimate of the noise, but without any assumptions about the origin of the signal itself. From this analysis, they could say that a transient, ``unmodeled'' signal was observed with a statistical significance greater than 4.6$\sigma$. The second method involved comparing the instrumental output (signal plus noise) with signals of merging black holes that were calculated using general relativity. From this so-called matched-filtering search, the researchers concluded that the significance of the observation was greater than 5.1$\sigma$.

The most exciting conclusions come from comparing the observed signal's amplitude and phase with numerical relativity predictions, which allows the LIGO researchers to estimate parameters describing the gravitational-wave source. The waveform is consistent with a black hole binary system whose component masses are 36 and 29 times the mass of the Sun. These stellar-mass black holes -- so named because they likely formed from collapsing stars -- are the largest of their kind to have been observed. Moreover, no binary system other than black holes can have component masses large enough to explain the observed signal. (The most plausible competitors would be two neutron stars, or a black hole and a neutron star.) The binary is approximately 1.3 billion light years from Earth, or equivalently, at a luminosity distance of 400 megaparsecs (redshift of $z\sim 0.1$). The researchers estimate that about $4.6\%$ of the binary's energy was radiated in gravitational waves, leading to a rotating black hole remnant with mass 62 times the mass of the Sun and dimensionless spin of 0.67.
From the signal, the researchers were also able to perform two consistency tests of general relativity and put a bound on the mass of the graviton -- the hypothetical quantum particle that mediates gravity. 
In the first test, they used general relativity to estimate the mass and spin of the black hole remnant from an ``early inspiral'' segment of the signal and again from a ``post-inspiral'' segment. These two different ways of determining the mass and spin yielded similar values. 
The second test was to analyze the phase of the wave generated by the black holes as they spiraled inward towards one another. This phase can be written as a series expansion in $v/c$, where $v$ is the speed of the orbiting black holes, and the authors verified that the coefficients of this expansion were consistent with the predictions of general relativity. By assuming that a graviton with mass would modify the phase of the waves, they determined an upper bound on the particle's mass of $1.2\times 10^{-22}$~eV/c$^2$, improving the bounds from measurements in our Solar System and from observations of binary pulsars. These findings will be discussed in detail in later papers.

In physics, we live and breathe for discoveries like the one reported by LIGO, but the best is yet to come. As Kip Thorne recently said in a \href{http://www.bbc.com/news/science-environment-34298363}{\underline{BBC interview}}, recording a gravitational wave for the first time was never LIGO's main goal. The motivation was always to open a new window onto the Universe.

Gravitational wave detection will allow new and more precise measurements of astrophysical sources. For example, the spins of two merging black holes hold clues to their formation mechanism. Although Advanced LIGO wasn't able to measure the magnitude of these spins very accurately, better measurements might be possible with improved models of the signal, better data analysis techniques, or more sensitive detectors. Once Advanced LIGO reaches design sensitivity, it should be capable of detecting binaries like the one that produced GW150914 with 3 times its current signal-to-noise ratio, allowing more accurate determinations of source parameters such as mass and spin.

The upcoming network of Earth-based detectors, comprising Advanced Virgo, KAGRA in Japan, and possibly a third LIGO detector in India, will help scientists determine the locations of sources in the sky. This would tell us where to aim ``traditional'' telescopes that collect electromagnetic radiation or neutrinos. Combining observational tools in this way would be the basis for a new research field, sometimes referred to as ``multimessenger astronomy''~\cite{Aasi:2013wya}. Soon we will also collect the first results from \href{https://lisapathfinder.org/}{\underline{LISA Pathfinder}}, a spacecraft experiment serving as a testbed for \href{https://www.elisascience.org/}{\underline{eLISA}}, a space-based interferometer. eLISA will enable us to peer deeper into the cosmos than ground-based detectors, allowing studies of the formation of more massive black holes and investigations of the strong-field behavior of gravity at cosmological distances~\cite{AmaroSeoane:2012km}.

With Advanced LIGO's result, we are entering the dawn of the age of gravitational wave astronomy: with this new tool, it is as though we are able to hear, when before we could only see. It is very significant that the first ``sound'' picked up by Advanced LIGO came from the merger of two black holes. These are objects we can't see with electromagnetic radiation. The implications of gravitational-wave astronomy for astrophysics in the near future are dazzling. Multiple detections will allow us to study how often black holes merge in the cosmos and to test astrophysical models that describe the formation of binary systems~\cite{Dominik:2014yma,Stevenson:2015bqa}. In this respect, it's encouraging to note that LIGO may have already detected a second event; a very preliminary analysis suggests that if this event proves to have an astrophysical origin, then it is likely to also be from a black hole binary system. The detection of strong signals will also allow physicists to test the so-called no-hair theorem, which says that a black hole's structure and dynamics depend only on its mass and spin~\cite{Berti:2009kk}. Observing gravitational waves from black holes might also tell us about the nature of gravity. Does gravity really behave as predicted by Einstein in the vicinity of black holes, where the fields are very strong? Can dark energy and the acceleration of the Universe be explained if we modify Einstein's gravity? We are only just beginning to answer these questions~\cite{Yunes:2013dva,Berti:2015itd}.


%


\end{document}